# 3D full-band, Atomistic Quantum transport in n-Si Junction less Nanowire field-effect transistors

Bhupesh Bishnoi[1] and Bahniman Ghosh[1,2]

[1]DEPARTMENT OF ELECTRICAL ENGINEERING, INDIAN INSTITUTE OF TECHNOLOGY, KANPUR, 208016, INDIA

Email: bbishnoi@iitk.ac.in

[2]MICROELECTRONICS RESEARCH CENTER, 10100, BURNET ROAD, BLDG. 160, UNIVERSITY OF TEXAS AT AUSTIN, AUSTIN, TX, 78758, USA

Email: bghosh@utexas.edu

## ABSTRACT

In the present work, we have investigated the quantum transport in n-Si junction less nanowire field-effect transistors using 3-D, full-band atomistic $sp^3d^5s^*$ spin-orbital coupled tight-binding method. We have investigated the $I_{DS}$–$V_{GS}$ transfer characteristics, $I_{DS}$–$V_{DS}$ output characteristics, ON-current, OFF-current leakage, subthreshold swing and energy-position resolved electron density spectrum $G_n$ (x, E) in n-Si junction less nanowire field-effect transistors. We also study $I_{DS}$–$V_{GS}$ transfer characteristics with variation of High-K gate materials. Quantum mechanical simulation is performed on the basis of Non-Equilibrium Green Function formalism to solve coupled Poisson-Schrödinger equation self-consistently for potentials and local density of state in n-Si junction less nanowire field-effect transistors.

## KEYWORDS:

# INTRODUCTION

Intensive research work is going on in Junction less Nanowire field-effect transistors as power-supply scaling below 0.5 V is possible in these devices and at low voltages Junction less Nanowire field-effect transistors can outperform aggressively scaled MOSFETs. [1] Hence, overall power consumption can be reduced in nanoelectronics integrated circuits by using Junction less Nanowire field-effect transistors. [2] In present scenario Junction less Nanowire field-effect transistors are promising candidate due to their steep subthreshold swing (SS), better ON to OFF current ratio and high drive current at low voltages operation. [3] In the ITRS 2011 roadmap Junction less Nanowire field-effect transistors operates on $V_{DD}$ lesser than 1 V, $I_{ON}$ current of 100 milli-amperes, $I_{ON}/I_{OFF} > 10^5$ and SS below 60 mV per decade. Recently, Junction less Nanowire field-effect transistors have been experimentally demonstrated in Ref. [1] and Effect of band-to-band tunneling on junctionless transistors and Junction less Nanowire Tunnel field-effect transistors are investigated in Ref. [4, 5] In this Article, we simulate and study $I_{DS}$–$V_{GS}$ transfer characteristics, $I_{DS}$–$V_{DS}$ output characteristics, ON-current, OFF-current leakage, subthreshold swing, energy-position resolved electron density spectrum $G_n (x, E)$ and variation of High-K gate materials on the scaling and design of n-Si junction less nanowire field-effect transistors of 2 nm thin channel structures with the gate lengths of 10 nm.

# DEVICE STRUCTURE

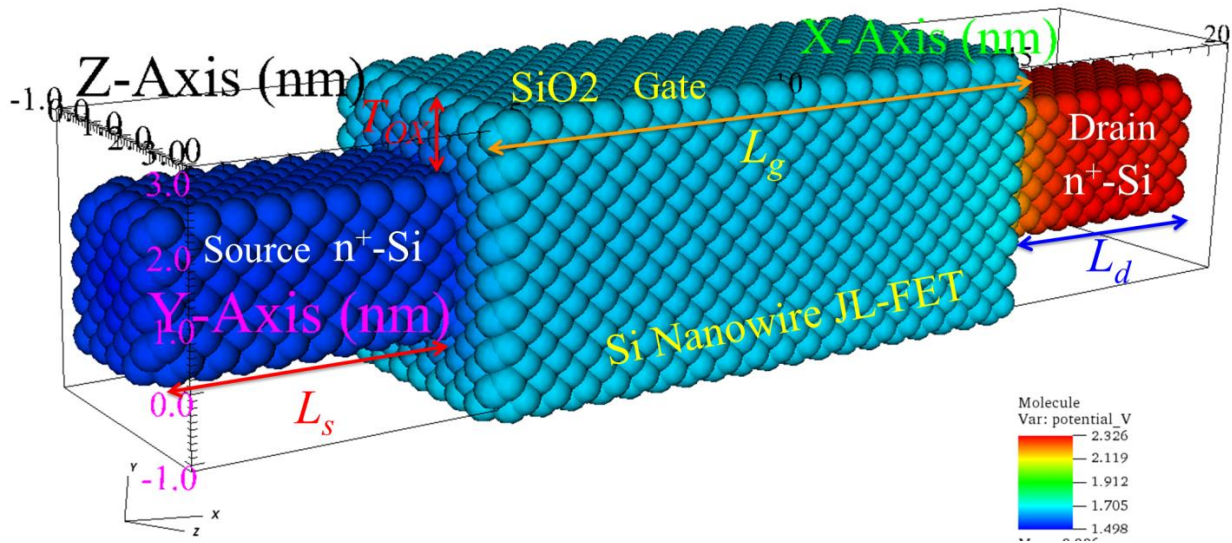

Fig.1. Atomic Structure and Geometry of n-Si-JL-NWFET

The n-Si-JL-NWFET has n-type Silicon channel of 10 nm with a doping density of $1\times10^{20}$ cm$^{-3}$ with source and drain region of 5 nm n-type Silicon with doping density of $1\times10^{20}$ cm$^{-3}$. Figure 1 shows the 3D atomistic structure of simulated n-Si-JL-NWFET device. As per the future ITRS 2012 roadmap for NWFET we started simulation at 10 nm gate length ($L_G$), 5 nm source length ($L_S$), 5 nm drain length ($L_D$), 2 nm thin channel structures and 1 nm SiO$_2$ gate thickness ($T_{OX}$). We also used High-K gate material of Al$_2$O$_3$ and HfO$_2$ for gate material variance study. We model source and drain as ohmic contact and gate is modeled as schottky contact with metal work function of 4.2 eV. Actual dimensions of Source is FEM grid: 2.172 x 2.172 x 4.888 [nm], Atom grid: 2.037 x 2.037 x 4.752[nm], Drain is FEM grid: 2.172 x 2.172 x 5.431[nm], Atom grid: 2.037 x 2.037 x 5.16[nm], Channel is FEM grid: 2.172 x 2.172 x 10.32 [nm], Atom grid: 2.037 x 2.037 x 10.18[nm] and Gate is FEM grid: 1.086 x 1.086 x 10.32[nm], Atom grid: 0.95 x 0.95 x 10.18[nm].

## SIMULATION APPROACH

We have investigated the quantum transport in n-Si junction less nanowire field-effect transistors using 3-D, full-band atomistic $sp^3d^5s^*$ spin-orbital coupled tight-binding method. [6-9] Quantum mechanical simulator based on atomistic $sp^3d^5s^*$ spin-orbital coupled tight-binding representation of the band structure solves Schrödinger and Poisson equations self-consistently. [10-13] Simulation also incorporates quantization effect due to narrow size effect and neglecting this will increase the band gap and cutoff distance to band edge is 0.05 eV. Carrier charge densities are self-consistently coupled to the calculation of electrostatic potential. Quantum mechanical simulation is performed on basis of Non-Equilibrium Green Function formalism to solve coupled Poisson-Schrödinger equation self-consistently including electron-phonon scattering matrix for potentials and local density of state in n-Si junction less nanowire field-effect transistors. [14-17] Transport direction is along <100> crystal axis in the channel and surface orientation is along (100). In the active region of device every atom is represented by a matrix and in the simulation Schrödinger equation is solved for 16068 active atoms. In the channel 12160 is number of finite elements with 2095 number of nodes taken for calculation. Atomistic tight binding calculation will give output as potential in volt and free charge in cm$^{-3}$ in the device and NEGF formalism will give output as current density (JE), charge density (NE), electron local density of states (n-LDOS), hole local density of states (p-LDOS) and output current. For solving the partial differential equations in the simulation PETSc toolkit is used and for the computation of eigenvalues and eigenvectors SLEPc library is used with krylov-schur algorithm for 70 numbers of eigenvalue with 7 digits of output precision. Gate dielectric layer is modeled as imaginary materials layer which has infinite bandgap as they separate the gate contacts and do not participate in transport calculation. Hence, in the Poisson equation they are characterized by their relative dielectric constant. [18-20]

RESULTS

### A. Energy-position resolved electron density spectrum $G_n (x, E)$

Figure 2 shows the energy-position resolved electron density spectrum $G_n (x, E)$ of the n-Si junction less nanowire field-effect transistors in the OFF-state ($V_{GS}$ 0 V to 0.2 V) at $V_{DS} = 0.8$ V. In the OFF-state biasing condition, channel is shut off for free charges. As $V_{GS}$ increases to 0.2 V charge carriers start penetrating in gate underneath channel region. These leakage charges are mainly due to carrier thermalization energy. The device is turned off by complete depletion of the channel region. The depletion is caused by the work function difference between the doped silicon and the gate of the nanowire. In comparison in conventional MOSFET device is turned off by reverse-biased p-n junction.

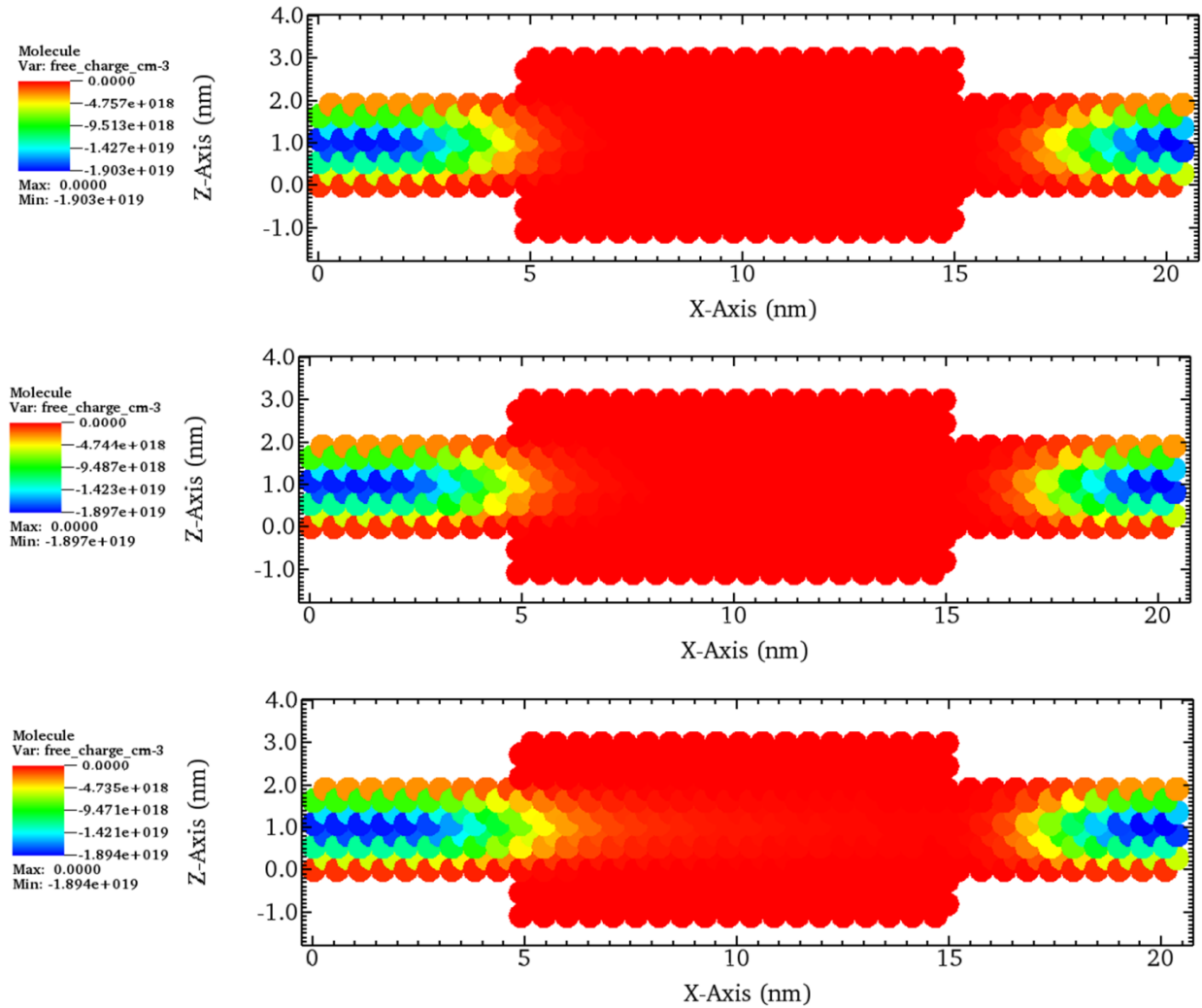

Fig.2. Energy-position resolved electron density spectrum $G_n(x, E)$ in the OFF-state ($V_{GS}$ 0 V to 0.2 V) of n-Si-JL-NWFET

Figure 3 shows the energy-position resolved electron density spectrum $G_n(x, E)$ of the n-Si junction less nanowire field-effect transistors ($V_{GS}$ ranging from 0.3 V to 0.5 V) at $V_{DS} = 0.8$ V. In this biasing range, channel starts forming from free charges. As $V_{GS}$ increases to 0.5 V charge carriers form channel underneath the gate. Gate starts modulating the position of the channel barrier and channel conduction band is pulled down below the source quasi Fermi level to increase the source injection as seen in Figure 3. The energy-position resolved electron density spectrum $G_n(x, E)$ as shown on log scale shows the occupation of LDOS $(x, E)$ by the respective source and drain contact Fermi reservoirs at $V_{DS} = 0.8$ V.

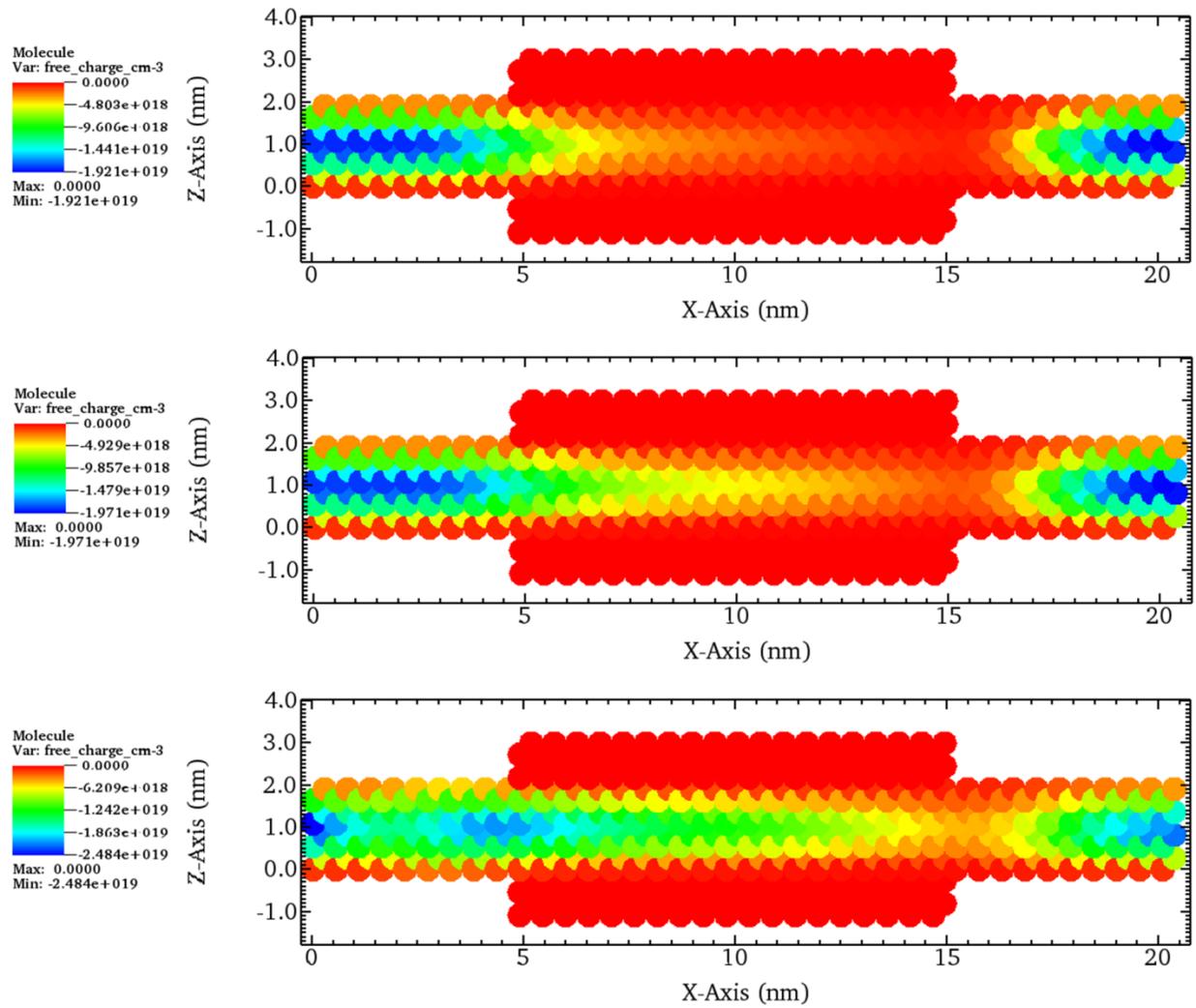

Fig.3. Energy-position resolved electron density spectrum $G_n(x, E)$ in the ($V_{GS}$ 0.3 V to 0.5 V) of n-Si-JL-NWFET

Figure 4 shows the energy-position resolved electron density spectrum $G_n(x, E)$ of the n-Si junction less nanowire field-effect transistors in the ON-state ($V_{GS}$ ranging from 0.6 V to 0.8 V) at $V_{DS} = 0.8$ V. As $V_{GS}$ increases to 0.8 V charge carriers form full channel underneath the gate. Gate modulates the position of the channel barrier and channel conduction band below the source quasi Fermi level and forms full channel. The energy-position resolved electron density spectrum $G_n(x, E)$ as shown on log scale shows the occupation of LDOS $(x, E)$ by the respective source and drain contact Fermi reservoirs at $V_{DS} = 0.8$ V.

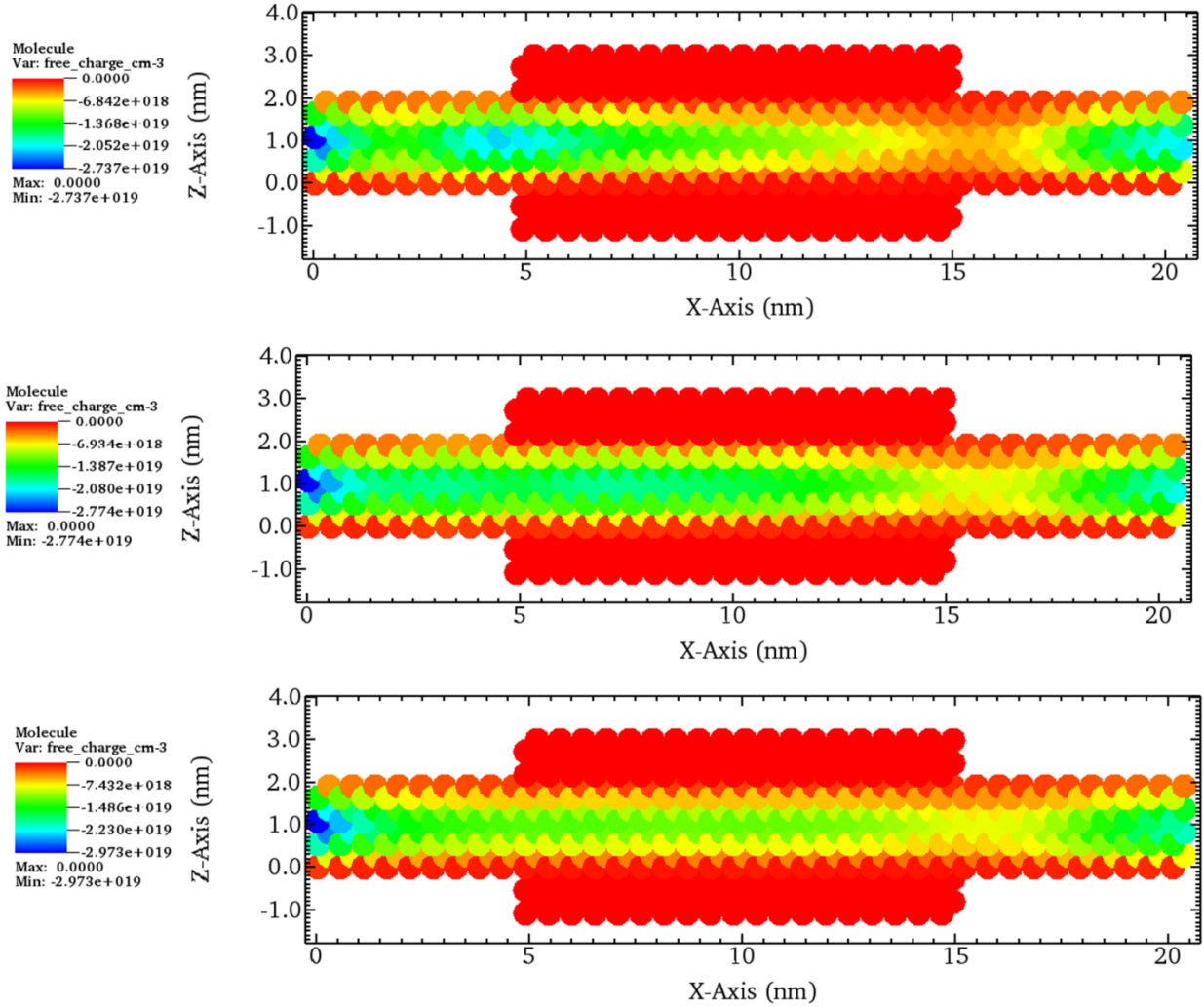

Fig.4. Energy-position resolved electron density spectrum $G_n(x, E)$ in the ON-state ($V_{GS}$ 0.6 V to 0.8 V) of n-Si-JL-NWFET

*B. Current-Voltage characteristics*

Figure 5 shows the $I_{DS}$-$V_{GS}$ transfer characteristics of n-Si junction less nanowire field-effect transistors at $V_{DS}$ of 0.8 V and 0.1 V. In the ON-state condition of $V_{DS}$ of 0.8 V on applying gate voltage $V_{GS}$ of 0.6 V, an $I_{ON}$ current of $1.05 \times 10^4$ µA/µm, an $I_{ON}/I_{OFF}$ ratio of ~$10^5$ with subthreshold swing of about 52.17 mV/decade are obtained. In the ON-state for gate voltage $V_{GS}$ higher than 0.6 V drain current saturates and transistor has high output resistance. $I_{OFF}$ current at $V_{GS}$ equal to 0 V is almost equal in both cases of $V_{DS}$ 0.8 V and 0.1 V which indicates the gate has effectively shut off the channel.

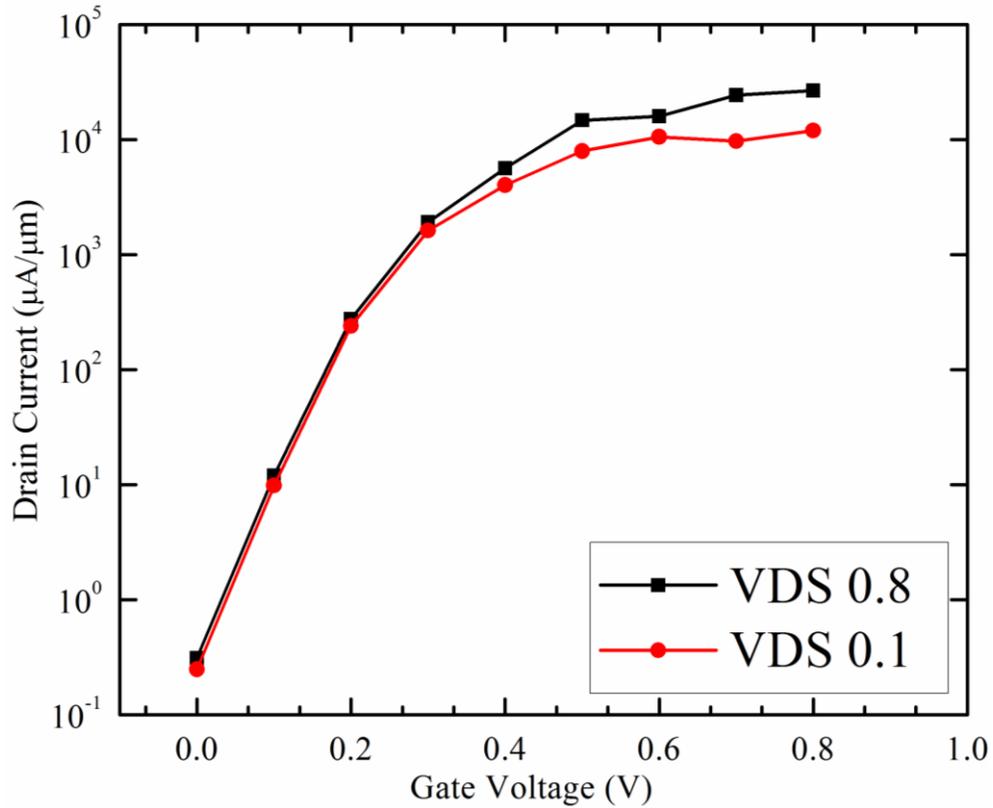

Fig.5. $I_{DS}$–$V_{GS}$ transfer characteristics at $V_{DS}$ = 0.1 V and $V_{DS}$ = 0.8 V in the n-Si-JL-NWFET

Figure 6 shows the $I_{DS}$–$V_{DS}$ output characteristics of n-Si junction less nanowire field-effect transistors with $V_{GS}$ variation of 0V, 0.2 V, 0.4 V, 0.6 V, and 0.8 V. For more than 0.1 V of $V_{DS}$ the drain current saturates for all values of $V_{GS}$ and drain current becomes almost constant at $V_{GS}$ equal 0.6 V and above. For $V_{GS}$ of 0 V the drain current is extinguished and thermionic emission governs the leakage current. At $V_{DS}$ larger than 0.1 V the drain current saturates for all the $V_{GS}$ ranging from 0.2 V to 0.8 V.

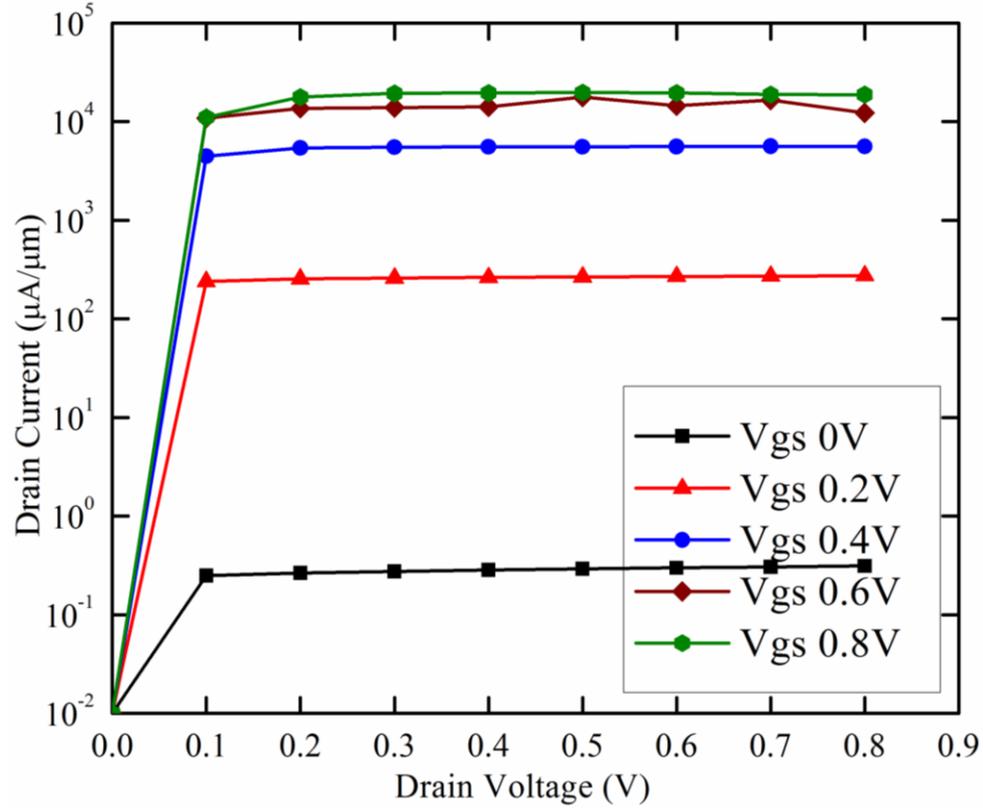

Fig.6. $I_{DS}$–$V_{DS}$ output characteristics with Gate voltage variation $V_{GS}$ in the n-Si-JL-NWFET

We also study the variation of different gate oxide material for n-Si junction less nanowire field-effect transistors as shown in the figure 7. For $Al_2O_3$ High-K Gate material subthreshold swing reduces to 37.5 mV/decade and for $HfO_2$ High-K Gate material subthreshold swing is 32.6 mV/decade.

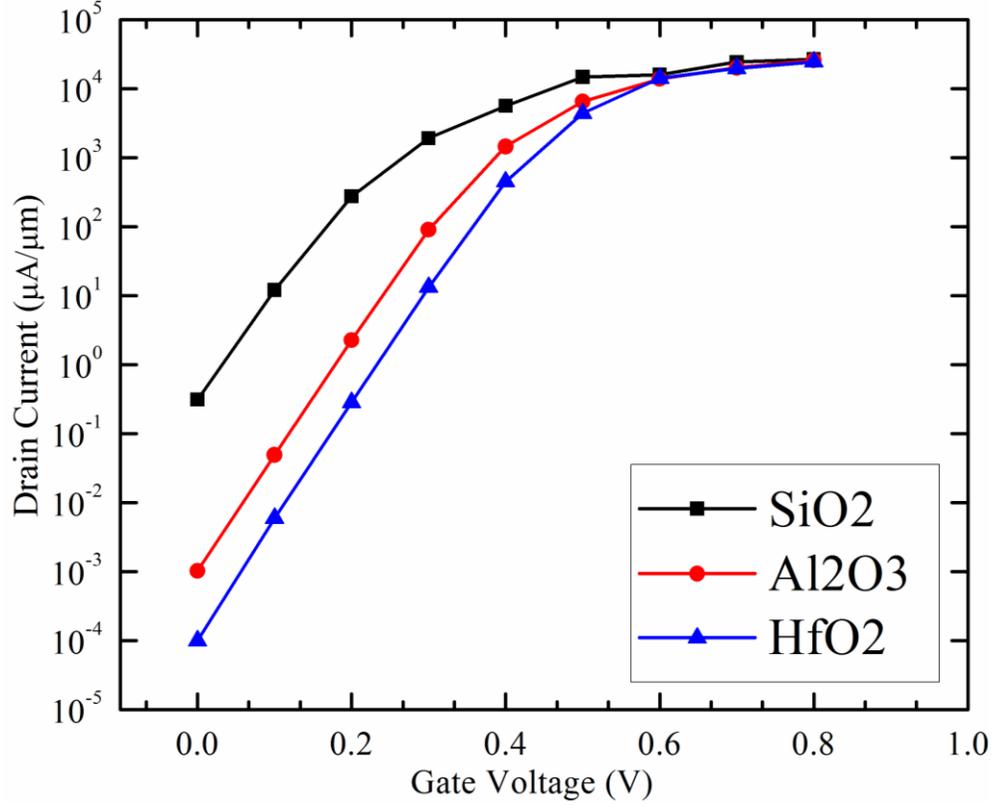

Fig.7. $I_{DS}$–$V_{GS}$ transfer characteristics at $V_{DS}$ = 0.8 V in the n-Si-JL-NWFET with variation in Gate Oxide Material

## CONCLUSION

We have investigated the quantum transport in n-Si junction less nanowire field-effect transistors of 10 nm channel length with a channel doping density of $1\times10^{15}$ cm$^{-3}$ with source and drain region of 5 nm n-type Silicon with doping density of $1\times10^{20}$ cm$^{-3}$. We used quantum mechanical simulator based on atomistic $sp^3d^5s^*$ spin-orbital coupled tight-binding representation of the band structure to solve Schrödinger and Poisson equations self-consistently in the NEGF formalism. We have investigated the $I_{DS}$–$V_{GS}$ transfer characteristics, $I_{DS}$–$V_{DS}$ output characteristics, ON-current, OFF-current leakage, subthreshold swing and energy-position resolved electron density spectrum $G_n$ (x, E) in n-Si junction less nanowire field-effect transistors. In the ON-state condition of $V_{DS}$ of 0.8 V on applying gate voltage $V_{GS}$ of 0.6 V, an $I_{ON}$ current of $1.05\times10^4$ µA/µm, an $I_{ON}/I_{OFF}$ ratio of ~$10^5$ with subthreshold swing of about 52.17 mV/decade are obtained. We also study $I_{DS}$–$V_{GS}$ transfer characteristics with variation of High-K gate materials. These simulation results suggest that it is worthy to experimentally demonstrate and investigate in more detail, the performance of n-Si junction less nanowire field-effect transistors for critical design parameter optimization and as building block for future low power nano-electronics circuits.

## ACKNOWLEDGEMENT

The authors thank the Department of Science and Technology of the Government of India for partially funding this work.